\documentclass{article}

\usepackage{amsmath}
\usepackage{amsfonts}
\usepackage{graphicx}

\begin{document} 

\begin{center}
{\bf COOPERATION AND SURVIVING WITH LIMITED RESOURCES}

\bigskip

K. Malarz$^1$ and K. Ku\l akowski$^2$

\bigskip

{\it 
Department of Theoretical and Computational Physics, Faculty of Physics and Nuclear Techniques, University of Mining and Metallurgy (AGH)\\
al. Mickiewicza 30, PL-30059 Krak\'ow, Poland}\\
E-mail: {$^1$malarz@agh.edu.pl, $^2$kulakowski@novell.ftj.agh.edu.pl}
\end{center}

\begin{abstract}
A network of agents cooperate on a given area. Time evolution of their power is described within a set of nonlinear equations. The limitation of resources is introduced via the Verhulst term, equivalent to a global coupling. Each agent is fed by some other agents from his neighborhood. Two subsequent stages of the time evolution can be observed. Initially, the richness of everybody increases distinctly, but its distribution becomes wide. After some transient time, however, resources are exhausted. Richness of some agents falls to zero and they are eliminated. Cooperation becomes less effective, what leads to subsequent falls. Finally, small percent of agents survive in a steady state. We investigate, how the cooperation influences the rate of surviving. 
\end{abstract}

\section{Introduction}
	Physicists are happy.
Their resources --- amount of problems to solve --- is infinite.
It is not so, however, in almost all other professions; the numbers of car buyers, voters, butterflies to catch and girls to kiss for the first time are limited.
If one takes everything, others have to look for another hobby or job.
On the other hand, cooperation is a fingerprint of modern society.
All what we get except from our domestic gardens, we get from other people.
The aim of this work is to present numerical results of a set of equations, where the above facts are built in as assumptions.
The only way to increase one's power or richness or speed of getting resources, which are treated here as synonymous, is to profit work of somebody else.
Then, each agent $i$ has $M$ neighbours $\{j(i)\}$ who feed its with given speeds $J_{ij}$. 
The limitation of resources is taken into account as the global coupling via a nonlinear Verhulst-like term. 
These are first two r.h.s. terms of our basic equations \eqref{eq1}, and these terms do not depend on $i$. 
Third term is to include some action of an $i$-th agent. 
Namely, it selects from his neighbours the one most sensitive for this action, and enhances feeding from this particular neighbour.

	The problem belongs to the large class of models designed to describe a competition for resources \cite{bib1}.
Most frequently, however, some kind of dynamic equilibrium is considered, and only a few authors are interested in an ultimate catastrophe.
Example giving, although statistical physics provides tools for analysing stock market \cite{bib2}, the term ``bankruptcy'' is absent in physical journals (see \cite{liebreich} and \cite{aleksiejuk} for exceptions).
It seems worthwhile to take a glance on a collision of expanding society with the boundary of limited resources.

\section{Model}
	Time evolution of the individual agent power $P_i$ is described by differential equation:
\begin{equation}
\begin{split}
\dfrac{dP_i(t)}{dt}=\lambda_1\cdot\sum_{j=1}^M J_{ij}P_j(t)r_j(t)-\lambda_2\cdot\left[\dfrac{1}{N}\sum_{j=1}^NP_j(t)r_j(t)\right]^2\\
+\lambda_3\cdot\max_{1\le j \le M}\left[d_{ij}P_j(t)r_j(t)\right]
\end{split}
\label{eq1}
\end{equation}
Coupling constant $J_{ij}$ (speed of feeding) and $d_{ij}$ (sensitivity) are random positive reals normalized to unity and fixed during simulation, while $\lambda_1$, $\lambda_2$ and $\lambda_3$ describe intensities of the three terms.
The constants $r_i$ are equal to one for active agents and to zero in other case.
At the beginning, all agents are active, but once $P_i$ is negative for given $i$, $r_i$ is switched to zero and the $i$-th agent is eliminated from the game. 
We deal with a set of differential equations which are piecewisely continuous. 
At the moments of time $t^*$ when $P_i(t^*)=0$ for any up-to-now-active agent $i$, the equations are switched from one analytical solution to another one. 
In this sense, the formalism is equivalent to a coupled map lattice, but the number of equations changes in time. 
Similar approach was applied already in \cite{bib4}.
Note however that in our case, the maps are to be integrated numerically.
We consider two different neighbourhoods: (i) each individual has $M$ randomly chosen neighbours, or (ii) each individual has $M$ nearest geometrical neighbours. 
In the latter case, the agents form a one-dimensional chain with periodic boundary conditions. 
Note that the formed network of agents is a continuous analogue to the Kauffman model \cite{bib5}, designed for a simulation of genetic systems.
We solve \eqref{eq1} numerically and check how the kind of neighbourhood, the number of directly interacting agents $M$ and the strengths of interactions influence (i) time evolution of the average power (ii) and number of active agents after very long times.
The simulation is started with $N_0=10^4$ active agents ($r_i=1$ for $i=1,\cdots ,N_0$), each with randomly chosen initial power $P_i(t=0)$. 
The simulation takes $N_{\text{iter}}=10^4$ time steps, each $10^{-3}$ long, what guarantees the numerical stability.

\begin{figure}
\begin{center}
(a) 
\includegraphics[width=8cm,angle=-90]{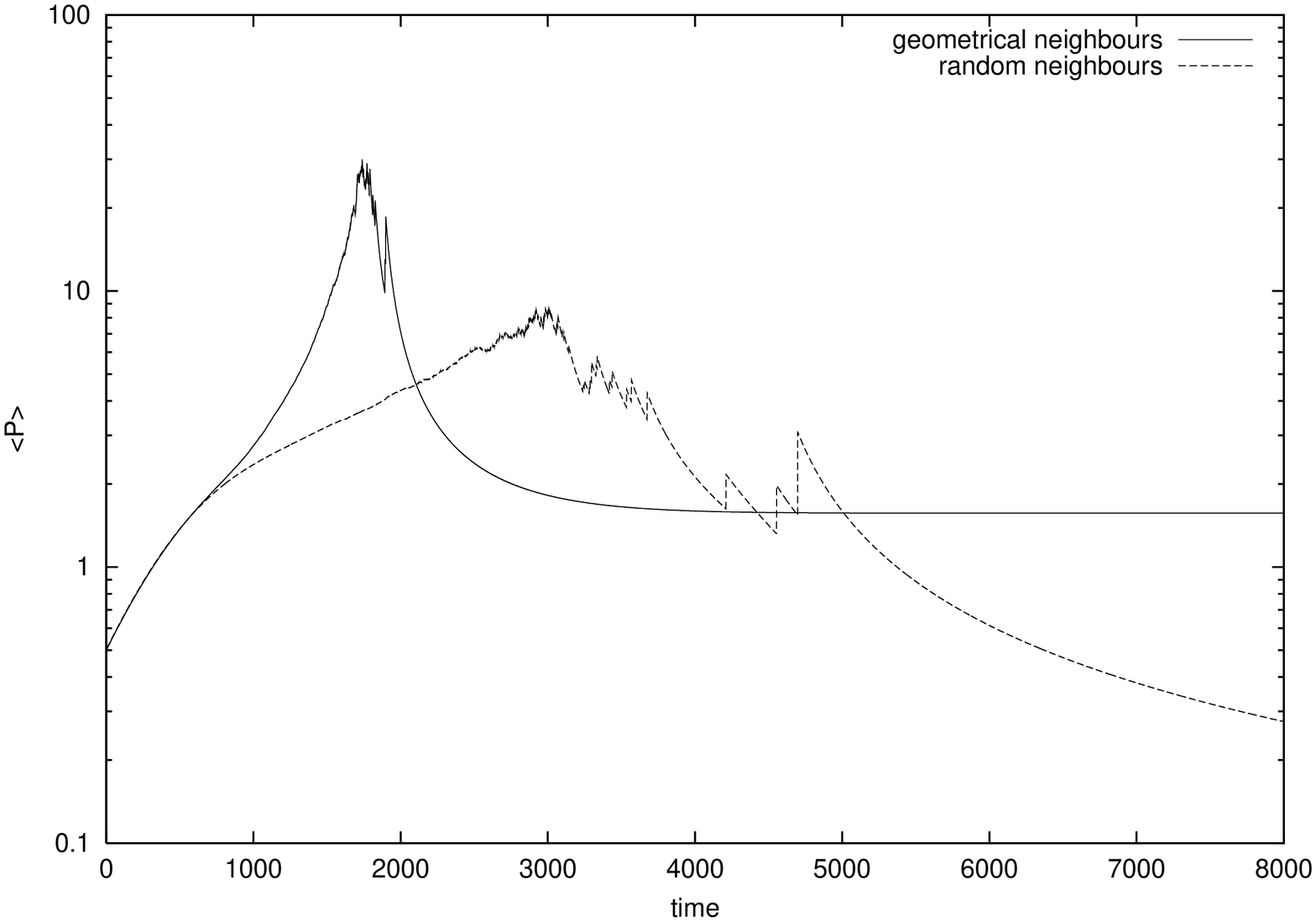}\\
(b) 
\includegraphics[width=8cm,angle=-90]{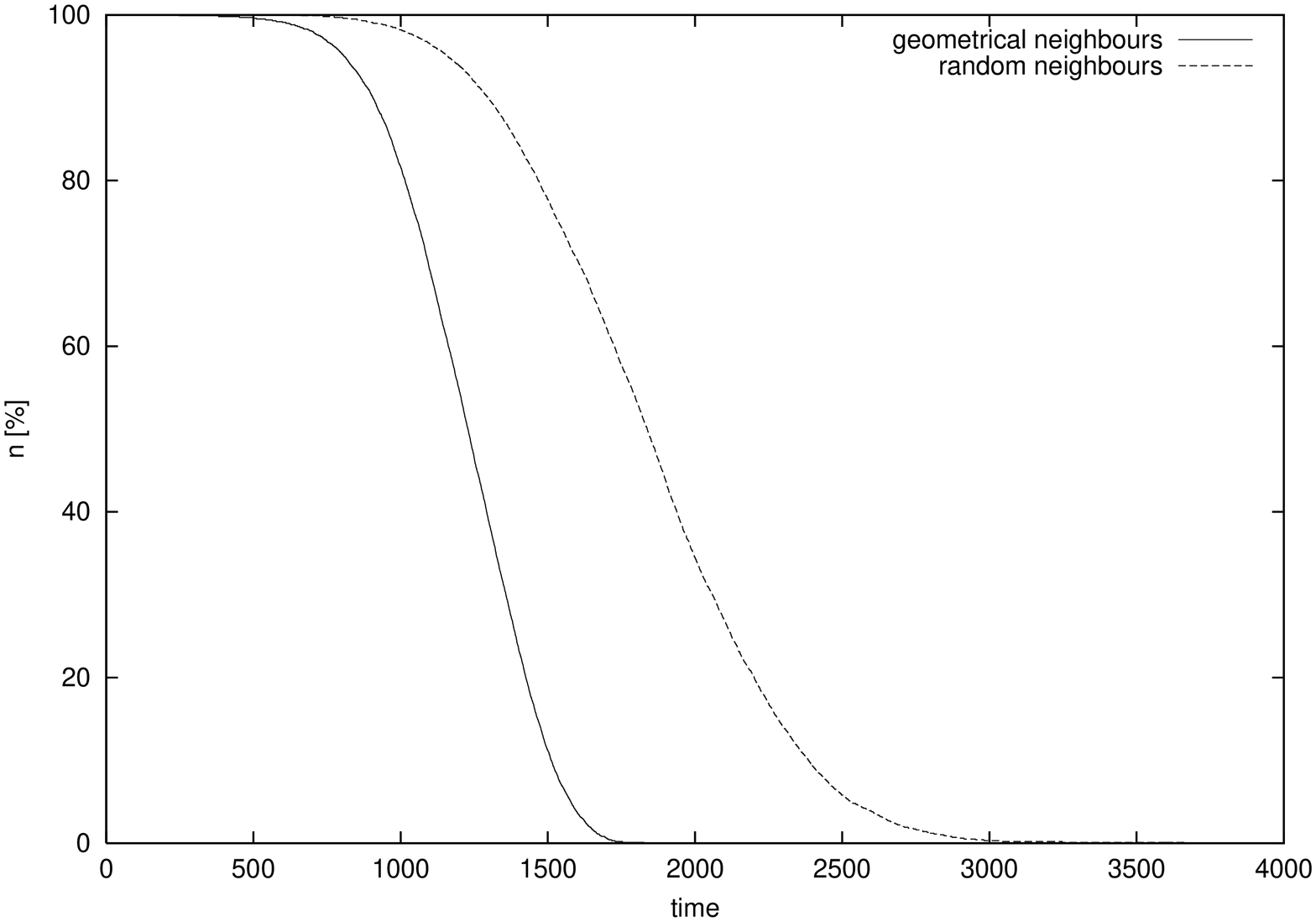}
\caption{Time evolution of (a) the average agent power $\langle P\rangle$ expressed in arbitrary units and (b) subsequent percentage $n$ of active agents for random (doted line) and geometrical (solid line) neighbours.}
\label{fig1}
\end{center}
\end{figure}

\section{Results and discussion}
	Two subsequent stages of the time evolution can be observed independently of kind of neighbourhood (Fig. \ref{fig1}).
Initially, the average power increases distinctly, but its distribution becomes wide.
After some transient time, however, resources are exhausted.
Richness of some agents falls to zero and they are eliminated.
Cooperation becomes less effective, what leads to subsequent falls.
Finally, small percent of agents survive in a steady state.
Distinct differences are found between the cases of geometrical and random neighbourhood (Fig. \ref{fig1}).
For the case of random neighbours, usually only one agent survives, $i=\text{ult}$. 
This simplifies Eq. \eqref{eq1} to
\begin{equation}
\dfrac{dP_{\text{ult}}(t)}{dt}=-\lambda_2\cdot P_{\text{ult}}^2(t)
\label{eq2}
\end{equation}
what gives subsequent power decrease of the last agent as $1/t$. 
If two or more cooperating agents survive, we are faced with a set of nonlinear equations. 
Analytically, a simplified case can be considered, when $J_{ij}=J$, $d_{ij}=d$. 
Then, the average power tends to a positive stable fixed point.
Numerical results suggest, that this is the rule also in the general case (Fig. \ref{fig1}). 
The marked difference between the results for random and geometrical neighborhoods is an illustration of the old truth {\em do ut des}.
In other words, it is better to help friends which can reward than to people randomly selected in the street.
A group of agents feeding each other can survive, if they spend resources moderately enough.
A kind of an equilibrium with a given environment seems to be possible for the geometrical neighborhood, as long as the Verhulst term is compensated by the remaining terms in Eq. \eqref{eq1}.
\begin{table}
\begin{center}
\begin{tabular}{ccccccccccc}
\hline
$\lambda_1$  $\lambda_3$
	& 0.5 & 1  & 1.5 & 2  & 2.5 & 3  & 3.5 & 4  & 4.5 & 5  \\
\hline
0.5 	& 78  & 63 & 61  & 59 & 58  & 55 & 48  & 52 & 49  & 48 \\
1 	& 93  & 77 & 71  & 65 & 60  & 60 & 58  & 54 & 54  & 52 \\
1.5 	& 96  & 84 & 76  & 73 & 67  & 65 & 62  & 59 & 57  & 44 \\
2 	& 96  & 91 & 85  & 78 & 71  & 70 & 65  & 63 & 60  & 37 \\
2.5 	& 97  & 94 & 89  & 82 & 75  & 72 & 70  & 65 & 64  & 59 \\
3 	& 97  & 95 & 92  & 86 & 81  & 76 & 74  & 69 & 67  & 66 \\
3.5 	& 97  & 96 & 92  & 90 & 86  & 79 & 76  & 73 & 70  & 68 \\
4 	& 97  & 97 & 95  & 91 & 87  & 89 & 78  & 75 & 71  & 72 \\
4.5 	& 97  & 96 & 95  & 92 & 89  & 86 & 81  & 78 & 78  & 75 \\
5 	& 97  & 96 & 96  & 93 & 92  & 88 & 83  & 81 & 80  & 73 \\
\hline
\end{tabular}
\caption{The percentage of the society's successes for geometrical neighbourhood and various sets of $(\lambda_1, \lambda_3)$. 
$M=4$, $\lambda_2=1$, $N_0=10^4$, $N_{\text{iter}}=10^4$, $N_{\text{run}}=1000$.}
\label{tab1}
\end{center}
\end{table}

\begin{table}
\begin{center}
\begin{tabular}{ccccccccccc}
\hline
$\lambda_1$ $\lambda_3$
	& 0.5  & 1    & 1.5  & 2    & 2.5  & 3    & 3.5  & 4    & 4.5  & 5   \\
\hline
0.5 	& 0.73 & 0.87 & 1.23 & 1.58 & 1.73 & 1.71 & 2.24 & 2.43 & 2.92 & 3.06\\
1 	& 1.30 & 1.50 & 1.56 & 1.97 & 2.15 & 1.99 & 2.84 & 2.92 & 3.56 & 3.71\\
1.5 	& 1.84 & 2.13 & 2.26 & 2.17 & 2.72 & 2.17 & 3.01 & 3.55 & 4.25 & 3.22\\
2 	& 2.43 & 2.54 & 2.70 & 2.97 & 2.74 & 3.31 & 3.22 & 4.07 & 4.29 & 4.92\\
2.5 	& 2.96 & 3.30 & 3.29 & 3.66 & 3.73 & 4.20 & 4.05 & 3.61 & 3.63 & 4.05\\
3 	& 3.47 & 3.63 & 4.11 & 3.76 & 4.42 & 4.09 & 4.06 & 4.06 & 4.34 & 4.54\\
3.5 	& 4.09 & 4.21 & 4.41 & 4.73 & 4.72 & 4.85 & 4.80 & 4.62 & 5.00 & 5.65\\
4 	& 4.63 & 4.68 & 5.02 & 5.10 & 5.77 & 4.91 & 5.52 & 5.26 & 5.88 & 5.89\\
4.5 	& 4.87 & 5.15 & 5.45 & 5.89 & 5.88 & 5.81 & 5.97 & 7.06 & 5.88 & 6.28\\
5 	& 5.43 & 5.68 & 5.87 & 6.19 & 6.48 & 6.20 & 6.96 & 6.58 & 6.53 & 6.76\\
\hline
\end{tabular}
\caption{Average power $\langle P\rangle$ (in arbitrary units) after $N_{\text{iter}}=10^4$ time steps averaged over $N_{\text{run}}=100$ independent runs for geometrical neighbourhood and various $(\lambda_1, \lambda_3)$.
$M=4$, $\lambda_2=1$, $N_0=10^4$.}
\label{tab2}
\end{center}
\end{table}

	The results given above refer to the case of the geometrical neighborhood.
Let us define as a success of a society the case when more than one cooperating agents survive.
Then, Tab. \ref{tab1} gives the percentage of successes for various sets of $(\lambda_1, \lambda_3)$.
It is astonishing (at least for us) that the roles of these two terms (first and third) in the r.h.s. of Eq. \eqref{eq1} differ so much.
Actually, both these terms are designed to increase the power of a given agent.
The difference is only that the third term is a kind of local optimization, while the first one is automatic.
It seems that this kind of dynamic reaction of a given agent is particularly relevant at the border of ``death'' (or ``bankruptcy'' or so), when $P_i$ is close to zero.
Average power in asymptotic time, presented in Tab. \ref{tab2}, does not show this effect.
There, both terms act in the same way and can be mutually replaced to get approximately the same result.

\section{Conclusions}
	We are somewhat surprised with the fury of crisis, which can be observed in Fig. \ref{fig1}.
Simultaneously, distinct but continuous progress of the average power is substituted by wild oscillations, and the number of agents is strongly reduced.
The effect arises abruptly, without any preceding warnings in the curve shape.
Soon, unavoidable elimination of almost all agents is observed for all applied sets of input parameters.
We imagine the model as a parabolic description of the process of  breaking of bonds and a destruction of a complex system.

	We feel that in this text, full of analogies, we are at the border of abuse of language.
We apologize for that.
However, we have found that it is particularly difficult to describe a transient process in precise terms of statistical physics, most of them designed for stationary processes.
Our process could be treated as stationary, if we allow the population of agents to be slowly reproduced, maybe with retaining some fruitful information, and the whole story is repeated many times.
This kind of investigation could link to a Penna-like model \cite{bib6} and to some kind of self-organization \cite{bib5} of the structure and strength of bonds between agents. 

\section*{Acknowledgment}
The authors are grateful to Dietrich Stauffer for criticism on manuscript and paying our attention to term ``bankruptcy'' and Ref. \cite{liebreich}.
The numerical calculations were carried out in ACK-CYFRONET-AGH.


\end{document}